\documentclass[]{aa}  

\usepackage{graphicx}
\usepackage{txfonts}
\begin{document}

   \title{Benchmarking the {\sl ab initio} hydrogen equation of state for the interior structure of Jupiter}

   \subtitle{}

   \author{
     S. Mazevet
          \inst{1,2}
     \and
     A. Licari\inst{1,3}
          \and
          F. Soubiran\inst{4}}

   \institute{IMCCE, Observatoire de Paris, Universit\'e PSL, CNRS, Sorbonne Universit\'e, Univ. Lille, Paris France\\
     \email{stephane.mazevet@obspm.fr}
     \and
     CEA-DAM-DIF, 91280 Bruy\`eres Le Chatels, France.\\
     \and
     CRAL, Ecole Normale Sup\'erieure de Lyon, 69364 Lyon Cedex 07, France.\\
     \and
      Laboratoire de G\'eologie de Lyon, Ecole Normale Sup\'erieure de Lyon, 69364 Lyon Cedex 07, France}

   \date{\today}

 
  \abstract{As Juno is presently measuring Jupiter's gravitational moments to unprecedented accuracy, models for the interior structure of the planet
        are putted to the test.  While equations of state based on
        first principles or {\sl ab initio} simulations have been available and used for the two most abundant elements constituting the envelope,
        hydrogen and helium, significant discrepancies remain regarding the predictions of the inner structure of Jupiter. The differences
        are severe enough to clutter the analysis of Juno's data and even cast doubts on the usefulness of these computationally expensive EOSs
        for the modeling of the interior of Jupiter and exoplanets at large}
      {Using our newly developed equations of state for hydrogen and helium, we asses the {\sl ab initio} equations of state
        currently available and establish their efficiency at predicting the interior structure of Jupiter in a two-layers model. We paid
        particular attention to the calculation of the total entropy for hydrogen that is required to
        calculate the convective H-He envelope but is a derived quantity from {\sl ab initio} simulations.}
      {The {\sl ab initio} equations of state used in this work are based on a parameterization of the
        {\sl ab initio} simulation points using a functional form of the Helmholtz free energy. It extends on our previous work recently published.
        Compared to previous {\sl ab initio} equations of state available, this latter approach provides an
        independent mean of calculating the entropy that was recently pointed out as deficient in some {\sl ab initio} results.}
      { By adjusting our free energy parameterization
          to reproduce previous {\sl ab initio} EOS behavior, we identify the source of the disagreement previously reported for the interior structure of Jupiter. We further point
          to area where care should be taken when building EOS for the modeling of giant planets. This concerns the interpolation between the {\sl ab initio} results
          and the physical models used to cover the low density range as well as the interpolation of the {\sl ab initio} simulation results at high densities. This
          sensitivity falls well within the uncertainties of the {\sl ab initio} simulations. This suggests that hydrogen EOS should be carefully benchmarked using a simple planetary
          model before being used in the more advanced planetary models
          needed to interpret the Juno data.  We finally provide an updated version of our {\sl ab initio} hydrogen EOS recently published.} 
        {}
   \keywords{equation of state, hydrogen, helium, Jupiter, planetary interiors, giant planets, exoplanets}

   \maketitle
%

   \section{Introduction}
   With the Juno spacecraft currently orbiting Jupiter, there is now a unique opportunity to constrain the planet inner structure
   using high-precision measurements of the gravitational field \citep{Folkner_2017,Bolton_etal17}. With the first orbits now completed and
   analyzed, several remarkable results are already reported. This includes the magnetic field properties of
   the planet, the depth of the atmospheric jet-stream estimated to extend $3000~$km below the surface \citep{Guillot_etal18}, and the suggestion
   that the core of the planet may be eroded and probably extending significantly outward in the envelope \citep{Wahl_etal17}. This latter result
   is potentially important to validate the core accretion scenario \citep{Pollack_etal96}, the formation model, and the time evolution of giant planets
   at large. It remains, however, cluttered by significant uncertainties regarding the predictions of the planet inner structure using various equations of state (EOS). 

   From the mid-90's, interior models of Jupiter mostly relied on the benchmark EOSs of  \citet{Saumon_1995} (SCVH). These EOSs provided the first
   comprehensive description of hydrogen and helium properties in the entire regime relevant to giant planets \citep{guillot_99}. These equations of state rest on
   a chemical description of the dense plasma and are obtained by minimizing the Helmholtz free energy of the system represented as a collection of atoms,
   molecules, ions, and electrons. This physically based model provides an estimation of the gradual dissociation and ionization of the hydrogen
   molecules taking place as the density increases along Jupiter's interior. Among its most notable features, it predicted that
   dissociation of hydrogen is a first order transition that occurs within Jupiter. This phase transition justifies a three layers model for Jupiter's interior
   consisting of a H-He envelope where hydrogen is neutral, an inner envelope where hydrogen turns metallic, and a solid core made of a mixture of water and
   silicates \citep{Stevenson_1977,Stevenson_1982}. This core is assumed to correspond to the primordial planetary embryo around which hydrogen and helium were accreted in the
   planetary nebula during the formation of the planet \citep{guillot_99,Pollack_etal96}. This understanding of the interior of Jupiter was completed by considering
   de-mixing of the hydrogen-helium mixture in the metallic envelope to account for the measurement of a sub-solar abundance of helium in the atmosphere measured
   by the Galileo probe \citep{galileo,guillot_99}. Demixing is also a convincing explanation of the strong depletion in neon observed by Galileo, as neon is
   more soluble in helium than in hydrogen \citep{WilsonMilitzer2010}. As mentioned earlier, the recent measurements by Juno support the existence of a diffuse core, blurring
   somewhat the three-layer pictures \citep{Wahl_etal17}. This is, however, beyond the scope of the present paper that mainly focuses on describing the H-He envelope
    and specifically on benchmarking the hydrogen EOS used.
   
   These predictions of Jupiter's inner structure based on the SCVH EOS  \citep{Saumon_1995} were putted into question in the early-2000s by shock measurements of
   hydrogen up to a few Mbars. These measurements indicated that along the principal Hugoniot, the dissociation and metalization of hydrogen do not coincide with
   a first order transition in the regime relevant to giant planet interior but happen rather continuously as pressure and temperature increase
   \citep{collins_1998,knudson_2001}. It further showed that hydrogen is not as compressible as predicted by the SCVH EOS. These findings immediately cast
   shadows on the resulting model of Jupiter's interior structure and triggered intense activities on both the experimental and theoretical sides, mostly
   focusing on the rate of dissociation of molecular hydrogen at planetary conditions \citep{saumon_2005}. 

   This lead to a new generation of EOS for hydrogen and helium based on density functional theory
   (DFT) \citep{lenosky_1997,militzer_2000,desjarlais_2003,holst_2008,Caillabet_Mazevet_Loubeyre11,hu_2011,Becker_etal14,Militzer_2013,Miguel_Guillot_Fayon16,chabrier_2019}.
   As demonstrated by their success at describing all the experimental data obtained so far, these {\sl ab initio} EOSs provide a description of the ionization and dissociation processes
   occurring along the principal shock Hugoniot and Jupiter's adiabat without adjustable parameters. Despite this now undisputed ability at providing an improved description
   of hydrogen and helium properties at planetary conditions \citep{knudson_2018}, the situation for the interior of Jupiter still remains cluttered. At the moment two different
   {\sl ab initio} EOSs published in the past ten years for the hydrogen-helium mixture are leading to significantly different
   predictions for the interior of Jupiter. The situation was recently summarized by \citet{Miguel_Guillot_Fayon16}. These two {\sl ab initio} EOSs are
   leading to drastically different predictions regarding the size of the core, the distribution of metallic elements within the envelope as well as the
   temperature profile within the planet \citep{Nettelmann_etal12,Militzer_2013,Miguel_Guillot_Fayon16}. These differences are significant enough to currently
   challenge our ability to correctly interpret the data from the Juno mission and, ultimately, to use this improved knowledge to validate formation models
   of giant planets.

   To resolve this issue, we recently developed equations of state for hydrogen, helium, and the associated H-He mixture based on density functional molecular dynamics
   simulations \citep{chabrier_2019,soubiran2012}.  Compared to previous {\sl ab initio} EOSs developed by \citet{Militzer_2013}  and  \citet{Becker_etal14},
   these latest EOSs rest on an independent mean of evaluating the Helmholtz free energy. This allows us to deduce the total
   entropy needed to model convective envelopes, that is in agreement with both the high-pressure melting properties as well as Monte Carlo simulations
    \citep{Caillabet_Mazevet_Loubeyre11}. Using these EOSs, we critically compare our results with previous predictions for both the entropy and the
   interior structure obtained for Jupiter. Following the work of  \citet{Miguel_Guillot_Fayon16,Miguel_2018}, we paid particular attention to the evaluation
   of the entropy for the case of hydrogen. To detangle the
     discrepancies reported so far, we adjust our free energy model to reproduce previous {\sl ab initio} simulation results. Lastly, we update our initial release of the
     hydrogen EOS \citep{chabrier_2019} to provide a more accurate fit of the {\sl ab initio} data in the thermodynamical
     regime relevant to the modeling of the interior structure of Jupiter.
   
   \section{Equation of state}
   The study reported here is based on a new set of EOSs for hydrogen and helium covering the complete thermodynamical range relevant
   to astrophysical modeling \citep{chabrier_2019,soubiran2012}. This density-temperature range extends significantly beyond the regime relevant to
   giant planets of the solar system to include hot exoplanets and brown dwarfs several times the size of Jupiter. This set of EOSs follows on the
   previous work of \cite{Caillabet_Mazevet_Loubeyre11} and complete {\sl ab initio} simulations data using results obtained from physical models
   for low and extreme densities as well as for high temperatures.  
   \begin{figure}
   \centering
   \includegraphics[width=6cm]{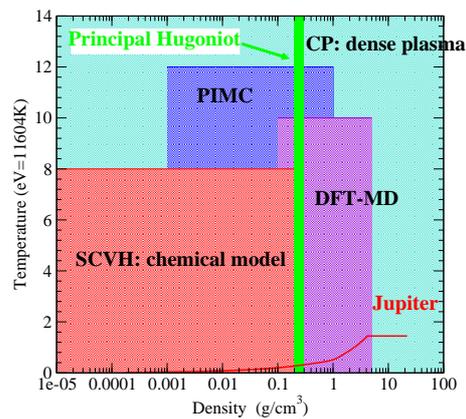}
   \caption{ Density-temperature diagram summarizing how the various methods are used to construct the \citet{chabrier_2019} hydrogen EOS relevant to astrophysical applications. (DFT-MD) density functional theory results of  \citet{Caillabet_Mazevet_Loubeyre11}, (PIMC) Path Integral Monte Carlo calculations of \citet{militzer_2000}, (CP) dense plasma model of \citet{chabrier_98}, (SCVH) chemical model of \citet{Saumon_1995}. A P-T version of the diagram can be found in \citet{chabrier_2019}.}
              \label{fig1}%
   \end{figure}
   Figure \ref{fig1} shows how the various methods are used to build a complete hydrogen EOS covering
   the complete thermodynamical range of interest for astrophysical applications. For hydrogen, density functional theory based molecular dynamics
   simulations (DFT-MD) can be used for densities down to 0.1-0.2 g/cm$^3$. Below this density range, the method becomes less efficient
   numerically. Standard DFT functionals, such as the commonly
   used PBE functional \cite{pbe}, also become less reliable. These functionals do not account for the van der Wall interactions that start to become relevant as the
   density decreases. The high-temperature limit of the method is more a practical one that steams from the number of Kohn-Sham orbitals that can be included in the simulation while
   keeping the overall simulation time tractable. 
   
   In the DFT-MD region, we used the parameterization of the {\sl ab initio} results provided by  \citet{Caillabet_Mazevet_Loubeyre11}.
   This consists in adjusting two physical models on the {\sl ab initio} results: a double-debye model for the solid phase and a
   one-component plasma model completed with a mass action law for dissociation to describe the liquid and plasma state. The input {\sl ab initio}
   data set used to obtain this parameterization includes the DFT-MD results of \citet{holst_2008} with additional calculations performed
   by  \citet{Caillabet_Mazevet_Loubeyre11}, coupled electron-ion Monte Carlo results of \citet{morales_2010}, which
   provides an evaluation of the Helmholtz free-energy in the liquid for temperatures ranging from $2000~$K to $10000~$K, and Path Integral Monte Carlo (PIMC)
   calculations of  \citet{militzer_2000}. Further comparisons with more recent PIMC calculations  \citep{hu_2011} can be found in \citet{chabrier_2019}.

   Figure \ref{fig1} shows that for hydrogen, we completed the dataset of \citet{Caillabet_Mazevet_Loubeyre11} with results of the
   dense plasma model (CP) \citep{chabrier_98} for densities above $5~$g/cm$^3$ and for temperatures above $10~$eV. At low densities, we used the
   SCVH  model \citep{Saumon_1995} to extend the {\sl ab initio} data set. The helium EOS used in the present work follows the
   same approach with the {\sl ab initio} data set based on DFT-MD simulations parameterized using a one-component plasma model and completed by both SCVH  and
   CP results at respectively low and high densities and CP results at high temperatures. Further details on the helium EOS can be found in \citet{soubiran2012} and has been discussed
   at length in \cite{chabrier_2019}.
   The challenge of building the overall EOS rests in assuring a smooth transition for the pressure, internal energy, and entropy as well as their derivative with respect to density
   and temperature. This complete set is needed for the calculation of the planetary interior structure. Figure \ref{fig1} indicates that for Jupiter, the transition that requires careful
   monitoring is between
   the low density SCVH results and the {\sl ab initio} ones. This transition between the two sets of calculations can be monitored by considering the principal Hugoniot.   
   \begin{figure}
   \centering
   \includegraphics[width=8cm]{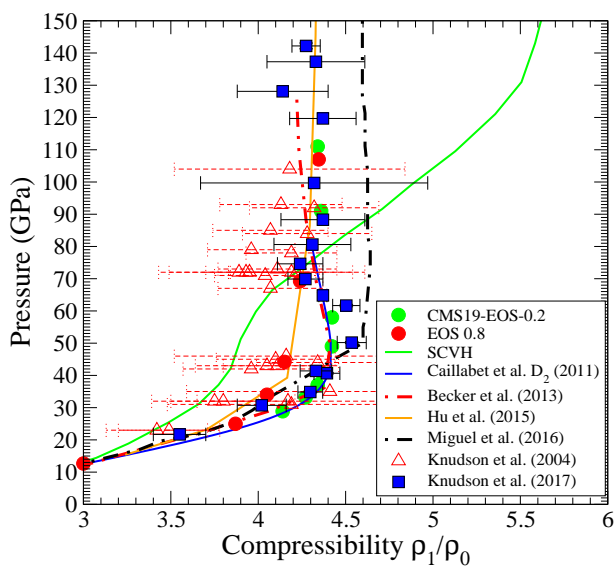}
   \caption{Comparison between the experimental data and the various theoretical predictions for the hydrogen principal Hugoniot. The initial state is taken as $\rho_0$=0.085g/cm$3$ and $T=20$K.  EOS-0.2 and EOS-0.8 are principal Hugoniot obtained with an interpolation range between the SCVH and {\sl ab initio} data extending from $0.05~$g/cm$^3$ to, respectively, $0.2~$g/cm$^3$ and $0.8~$g/cm$^3$. The former corresponds to the \cite{chabrier_2019} EOS and is thus labeled CMS19-EOS-0.2 in the figure.}
              \label{fig2}%
\end{figure}

   In figure \ref{fig2}, we compare the various theoretical predictions with the latest experimental data for the principal Hugoniot of hydrogen
    \citep{knudson_2004,knudson_2018}. We first notice that the latest analysis of the shock Hugoniot data by  \citet{knudson_2018} significantly reduced the
   error bars. It further indicates a maximum compressibility of $\rho_1/\rho_0=4.534$, which is slightly higher than previously reported \citep{knudson_2004}. When
   compared to the theoretical predictions, we first see, as reported previously, that the SCVH EOS misses the experimental data significantly. We further find
   that the results of  \citet{Miguel_Guillot_Fayon16} overestimates the maximum compressibility from $50~$GPa and beyond. The other predictions are slightly softer
   than the re-analyzed data set and stand just outside the error bars. The remaining difference in compressibility has been identified as the need for higher level
   functionals \cite{knudson_2018}. We further notice that the EOS recently published by \citet{chabrier_2019}, noted CMS19 from now on, agrees nicely with this revised version
   of the experimental data. As expected, it is in agreement with the {\sl ab initio} data \citep{Caillabet_Mazevet_Loubeyre11} used to build it.

   This revision of the experimental data has a noticeable impact for the modeling of the inner structure of Jupiter. The principal Hugoniot represents the density-temperature
   conditions at the edge of the interpolation region between the {\sl ab initio} data and the SCVH results. Reducing the error bars on the Hugoniot measurements
   adds a stronger constrain on the interpolation procedure in a density region where the two data sets do not coincide for either the internal energy or the entropy.
   The original shock data \citep{knudson_2004} allowed for a looser interpolation extending on a density region between $\rho=0.05$ and $0.8$g/cm$^3$. Figure \ref{fig2} shows that
   this leads to a principal Hugoniot, EOS-0.8, with a maximum compressibility reduced and compatible with the original shock data, the predictions of \citet{hu_2011},
   and \citet{kerley_2013} (not shown in figure \ref{fig2}).
   The re-analyzed shock data lead to a reduction of the density range over which the interpolation between the SCVH and the {\sl ab initio} data is performed, between
     $\rho=0.05$ and $0.2$g/cm$^3$ (labeled as CMS-19-EOS-0.2 in figure \ref{fig2}).

   \begin{figure}
   \centering
   \includegraphics[width=6.5cm]{fig2-a.eps}
   \includegraphics[width=6.5cm]{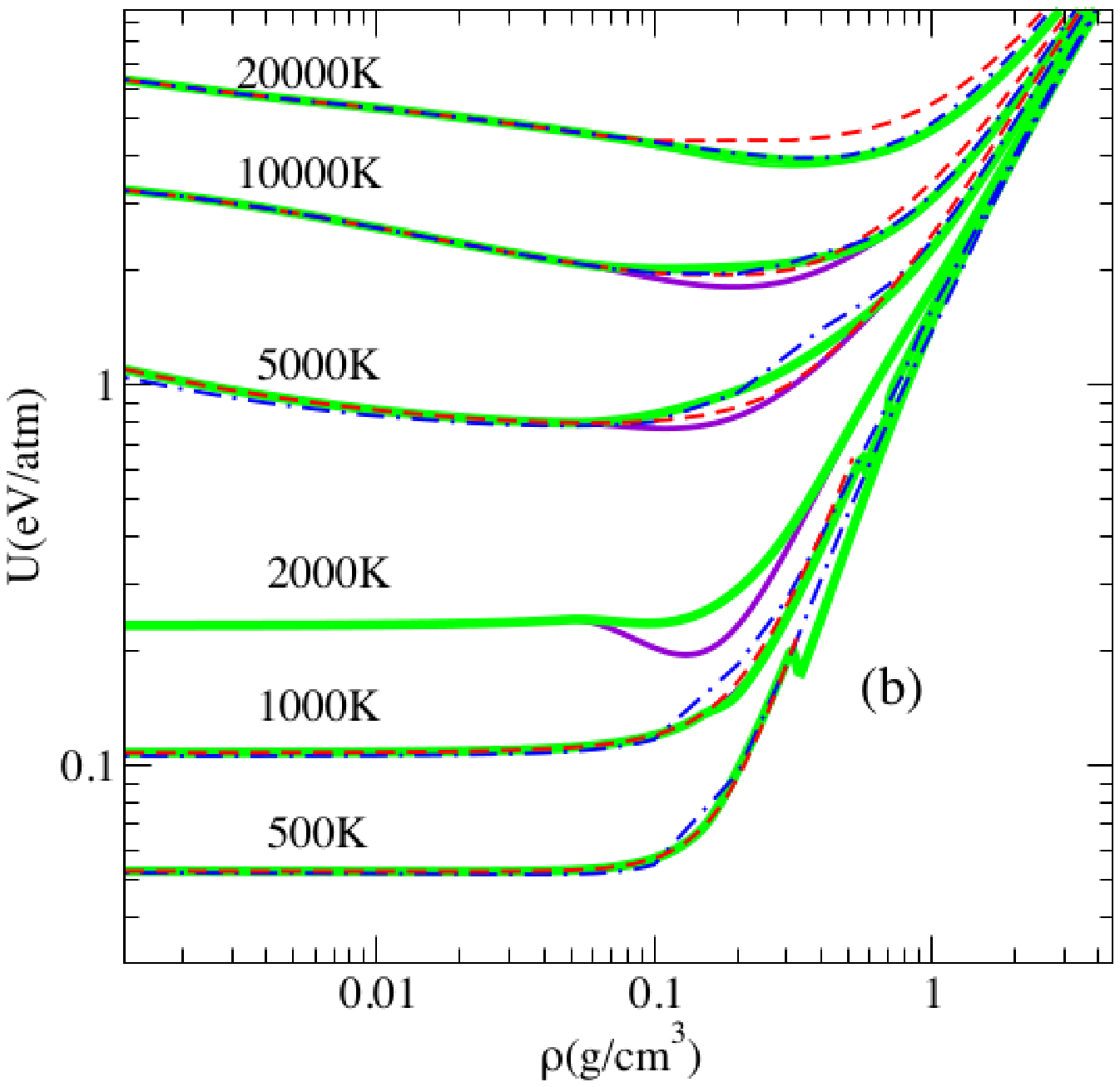}
   \caption{Comparison between the (a) pressure normalized to the density and (b) internal energy as given by various EOS for hydrogen. Same legend as in figure \ref{fig2}, MH-SCVH stands for the \citet{Militzer_2013} EOS as calculated by \citet{Miguel_Guillot_Fayon16}. }
              \label{fig3}%
   \end{figure}
   
   In figure \ref{fig3}, we compare the pressure and internal energy for hydrogen as given by our recently published EOS, CMS-EOS-0.2, with the latest
   EOS of  \citet{Becker_etal14}, the standard SCVH EOS \citep{Saumon_1995} and the \citet{Militzer_2013} EOS as calculated by
    \citet{Miguel_Guillot_Fayon16}, MH-SCVH. We see that differences between the SCVH result and all the {\sl ab initio} results are clearly visible for
    both the pressure and internal energy at densities above $0.1~$g/cm$^3$. This has been previously documented by several authors.
   
   Figure \ref{fig3} also shows that the overall agreement between the various {\sl ab initio} results is, at first glance, rather satisfactory with
   some differences above 0.1g/cm$^3$. As pointed out above, this density range corresponds to the interpolation region between the SCVH result and the
   {\sl ab initio} data for our newly developed EOS \citep{chabrier_2019}, the EOS of  \citet{Becker_etal14}, as well as the one from
   \cite{Miguel_Guillot_Fayon16}. We point out here that  \citet{Miguel_Guillot_Fayon16} extracted an EOS for pure hydrogen from the H-He EOS
   of  \citet{Militzer_2013} that includes the non-ideal contribution to the entropy of mixing. They further used the He SCVH EOS that differs from the {\sl ab initio}
   results \citep{soubiran2012}.
   Figure \ref{fig3}-a, where we display the pressure normalized to the density, shows that the interpolation appears rather consistent between the
   various EOSs when considering the pressure. In contrast, figure \ref{fig3}-b indicates noticeable differences in the transition region for the internal energy
   and for all the temperatures relevant to giant planets. The behavior of the $2000$ and $5000~$K isotherms obtained with the EOS-0.8 and \citet{Becker_etal14} EOS suggests
   that this is a delicate domain. \citet{Miguel_Guillot_Fayon16}  noticed this behavior for the internal energy and speculated that it propagates to the
     evaluation of the entropy. We confirm this result here and confirm that it may explain some of the differences in the temperature profile of Jupiter obtained using
     various {\sl ab initio} EOSs.

   \begin{figure}
   \centering
   \includegraphics[width=8cm]{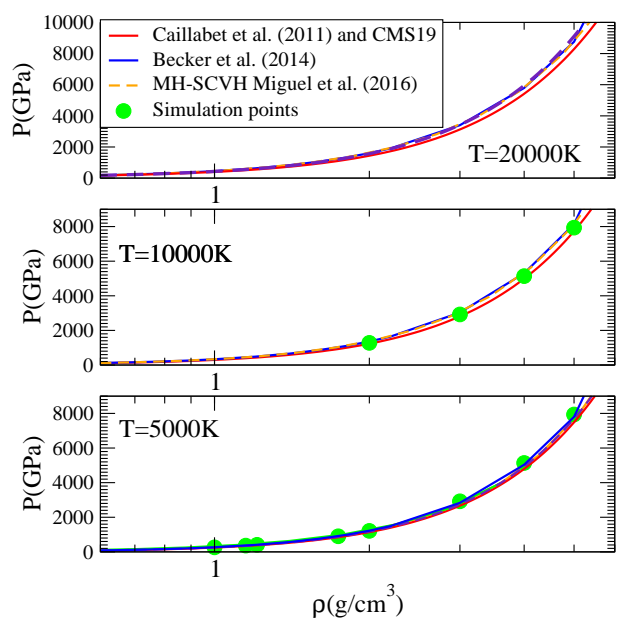}
   \caption{ Comparison between the initial {\sl ab initio} simulation points of \citet{Caillabet_Mazevet_Loubeyre11} and the various {\sl ab initio} EOS available at high-density.}
              \label{fig4}%
   \end{figure}
   
   To further compare the various {\sl ab initio} EOSs available for the modeling of Jupiter, we show in figure \ref{fig4} the high-density behavior along three isotherms
   particularly relevant for the interior structure of Jupiter. This density-temperature regime corresponds to the planet deep envelope and close to the core-envelope boundary
   where significant differences have been reported between the various {\sl ab initio} predictions. Figure \ref{fig4} shows some
   differences at high densities between the three EOSs for all the temperatures investigated. When comparing to the raw {\sl ab initio} data points, we see that the EOSs remain
   within 5\% of the simulation result. For a temperature of $T=10000~$K and a density of $\rho=4~$g/cm$^3$, the original  \citet{Caillabet_Mazevet_Loubeyre11} fit is
   2\% lower than the {\sl ab initio} data while both the  \citet{Miguel_Guillot_Fayon16} and  \citet{Becker_etal14} are 4\% higher. This difference worsen as
   temperature increases. At $T=20000~$K and for the same density range, the difference in pressure between the  \citet{Becker_etal14},  \citet{Miguel_Guillot_Fayon16},
   and  \citet{Caillabet_Mazevet_Loubeyre11} reaches 7\%. While this stays within the prescribed boundary for a fit adjusted over a broad density-temperature
   range, this difference is significant enough to require attention when considering the interior structure of Jupiter.  We recall here that the \citet{chabrier_2019} EOS is based on
   the \citet{Caillabet_Mazevet_Loubeyre11} parameterization.  

   We further point out that the {\sl ab initio} simulation points themselves have some uncertainty associated to them. Various simulation parameters such as the
   number of particles used in the simulation cell, the plane wave cut-off, the functional used, as well as the fluctuation naturally occurring within the simulation all bring
   a combined uncertainty of a few percents in the final pressure. In this context, a difference of a few percents between the different EOS based on {\sl ab initio}
   results can be expected. While this likely explains the difference between our EOS \citep{chabrier_2019} and the result of \citet{Becker_etal14}, we also mention that the
     hydrogen EOS extracted by \citet{Miguel_Guillot_Fayon16} from the H-He of \cite{Militzer_2013} adds an additional uncertainty by using the SCVH He EOS and by neglecting the non-ideal
   mixing contribution accounted for in the \citet{Militzer_2013} EOS. The former probably explains the difference we see here for the pressure between the hydrogen EOS extracted by
   \citet{Miguel_Guillot_Fayon16} and the {\sl ab initio} simulation points.

   To test how this uncertainty propagate for the inner structure of Jupiter, we adjusted the
   parameterization provided by \citet{Caillabet_Mazevet_Loubeyre11} and used in \citet{chabrier_2019} to reproduce precisely the {\sl ab initio} results and the pressures obtained
   by the two other {\sl ab initio} EOSs. This is obtained by writing $d_c(\rho)=d_0\rho$ for $\rho\geq 0.419$  and varying the $d_0$ parameter to a value
   of respectively, $d_0=0.0326858$, $0.123$, and $0.223$. The first value corresponds to the initial parameterization of \citet{Caillabet_Mazevet_Loubeyre11}
   used in \citet{chabrier_2019} and \citet{debras_2019}, the second matches exactly the {\sl ab initio} results, while the latest value reproduces the \citet{Becker_etal14}
   behavior at high densities.

\begin{figure}
   \centering
   \includegraphics[width=7cm]{fig4-a.eps}
   \includegraphics[width=7.5cm]{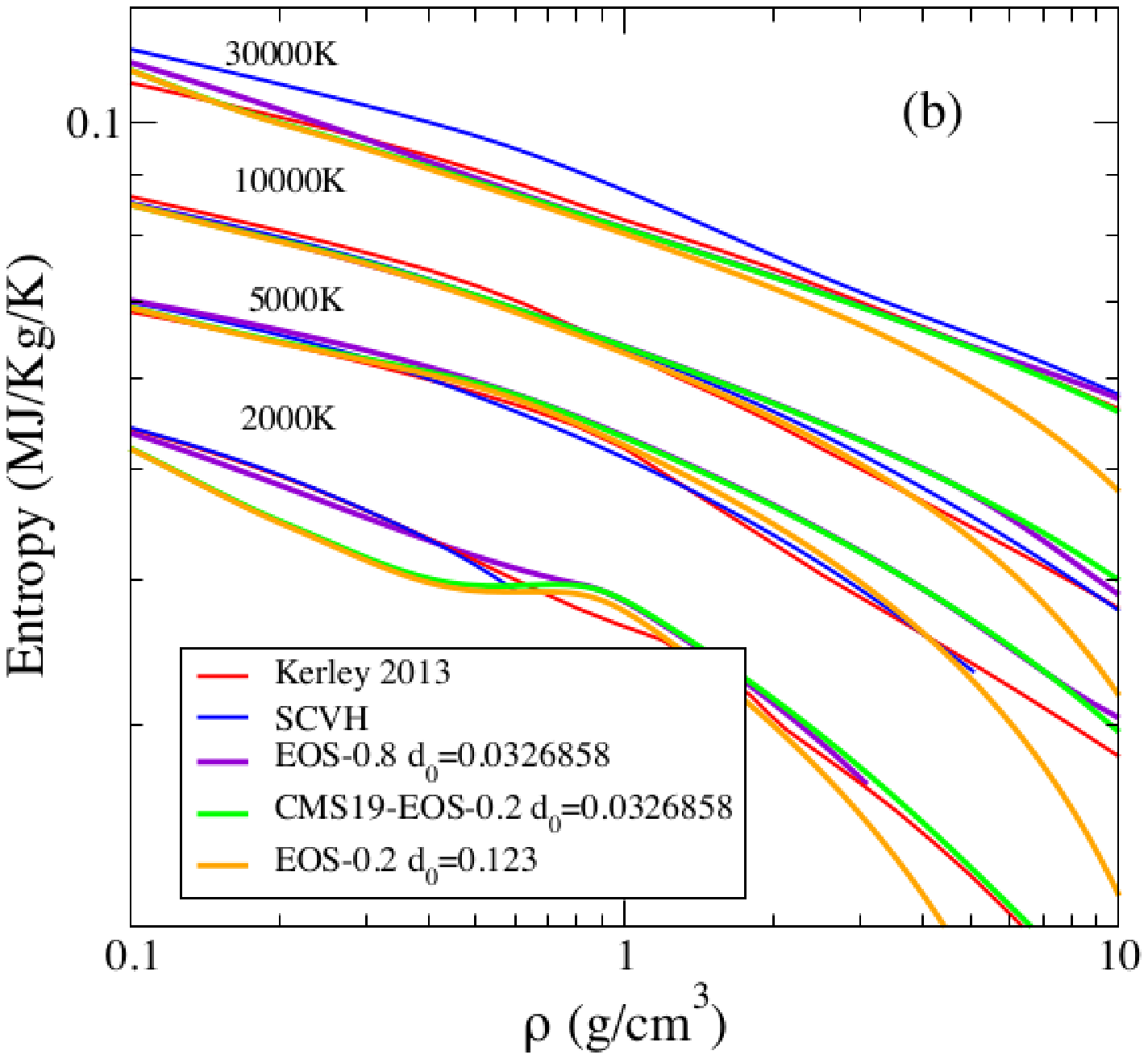}
   \caption{Variation of the hydrogen entropy as a function of density for isotherms relevant to Jupiter modeling. (a) Comparison between the various {\sl ab initio} predictions and the coupled electron-ion Monte Carlo result \citep{morales_2010}. (b) Comparison between the CMS-EOS-0.2 $d_0=0.0326858$) EOS for hydrogen, with results considering interpolation in a large density range, EOS-0.8 $d_0=0.0326858$, or adjusting the high pressure behavior, EOS-0.2 $d_0=0.132$, and with the SCVH predictions and another well used physically based EOS \citep{kerley_2013}.}
              \label{fig5}%
   \end{figure}

To address the issue of the evaluation of the entropy pointed out by \citet{Miguel_Guillot_Fayon16}, we show in figure \ref{fig5} the various {\sl ab initio} predictions for isotherms
representative of the interior of Jupiter. The labeling follows the one given by \citet{Miguel_Guillot_Fayon16} where the entropy deduced from the EOS of \citet{Becker_etal14} and corrected
for the energy are labeled respectively eos3b and eos3c. The entropy deduced from the H-He EOS of \citet{Militzer_2013} is labeled MH-SCVH. We see in figure \ref{fig5}-a that the predictions
are overall rather consistent and in rather good agreement with the coupled electron-ion Monte Carlo calculations of \citet{morales_2010}.  The latter is considered as a benchmark
  result for the entropy. As expected, the largest differences are seen
in the interpolation region for densities between $\rho=0.1~$g/cm$^3$ and $\rho=0.8~$g/cm$^3$. These differences reflect the variation that was observed for the principal Hugoniot
predictions shown in figure \ref{fig2}. We also note that once corrected \citep{Miguel_2018}, the high-temperature behavior initially noticed for the entropy deduced
from the EOS of \citet{Becker_etal14} using thermodynamics relations tends to disappear.  The results shown in figure \ref{fig5} suggest that the estimation of the entropy is rather
  consistent between the various models. The evaluation of the entropy itself is not likely the direct cause of the differences that are reported in the predictions for the interior of Jupiter.

This is confirmed in figure \ref{fig5}-b, where we compare the predictions of our EOS \citep{chabrier_2019} with the results given by the SCVH and Kerley \citep{kerley_2013} EOSs.
The latter is a refined version of the SCVH EOS that takes into account the experimental Z-pinch data of \citet{knudson_2004}. In the interpolation region, the difference  is
surprisingly only noticeable at the lowest and highest temperatures displayed. We also show in figure \ref{fig5}-b the effect of varying the interpolation region between
the SCVH and the {\sl ab initio} results from the density range extending from $0.05$ to $0.2$g/cm$^3$ to $0.05$ to $0.8$g/cm$^3$ while keeping $d_0=0.0326858$. These EOS are labeled as,
respectively, CMS19-EOS-0.2 and EOS-0.8. At $T=2000~$K, we see that a larger interpolation range provides a smoother transition between the SCVH and {\sl ab initio} data sets.
The EOS-0.8-$d_0=0.0326858$ result is almost indistinguishable from the SCVH result. We recall from the discussion of figure \ref{fig2} that this occurs at the expense of not
reproducing the re-analyzed shock data. As this also corresponds to the differences in the internal energy noticed between the various models, this tends to further suggest that the different
predictions for the internal energy only impact the entropy deduced in the low-temperature region of the isentrope. It is thus not likely the origin of the difference noted for the whole
adiabat of Jupiter.

 We also show in figure \ref{fig5}-b how the entropy changes  when the EOS is adjusted to reproduce the initial {\sl ab initio} calculations to less than 1\%; noted as EOS-0.2-$d_0=0.123$. We see that adjusting the pressure at high densities reproduces the variations noticed between the various {\sl ab initio} predictions at high densities. This suggests that
  the evaluation of the entropy itself, which varies in the various methods, is probably not a cause for concerns. The variation between the various predictions is mostly a propagation of
  the difference noticed for the energy and pressure in the interpolation between the {\sl ab initio} et SCVH results and, at higher densities, in the interpolation of the {\sl ab initio} results.

\section{Jupiter Inner Structure}

\begin{figure}
   \centering
   \includegraphics[width=8cm]{fig5-a.eps}
   \includegraphics[width=8cm]{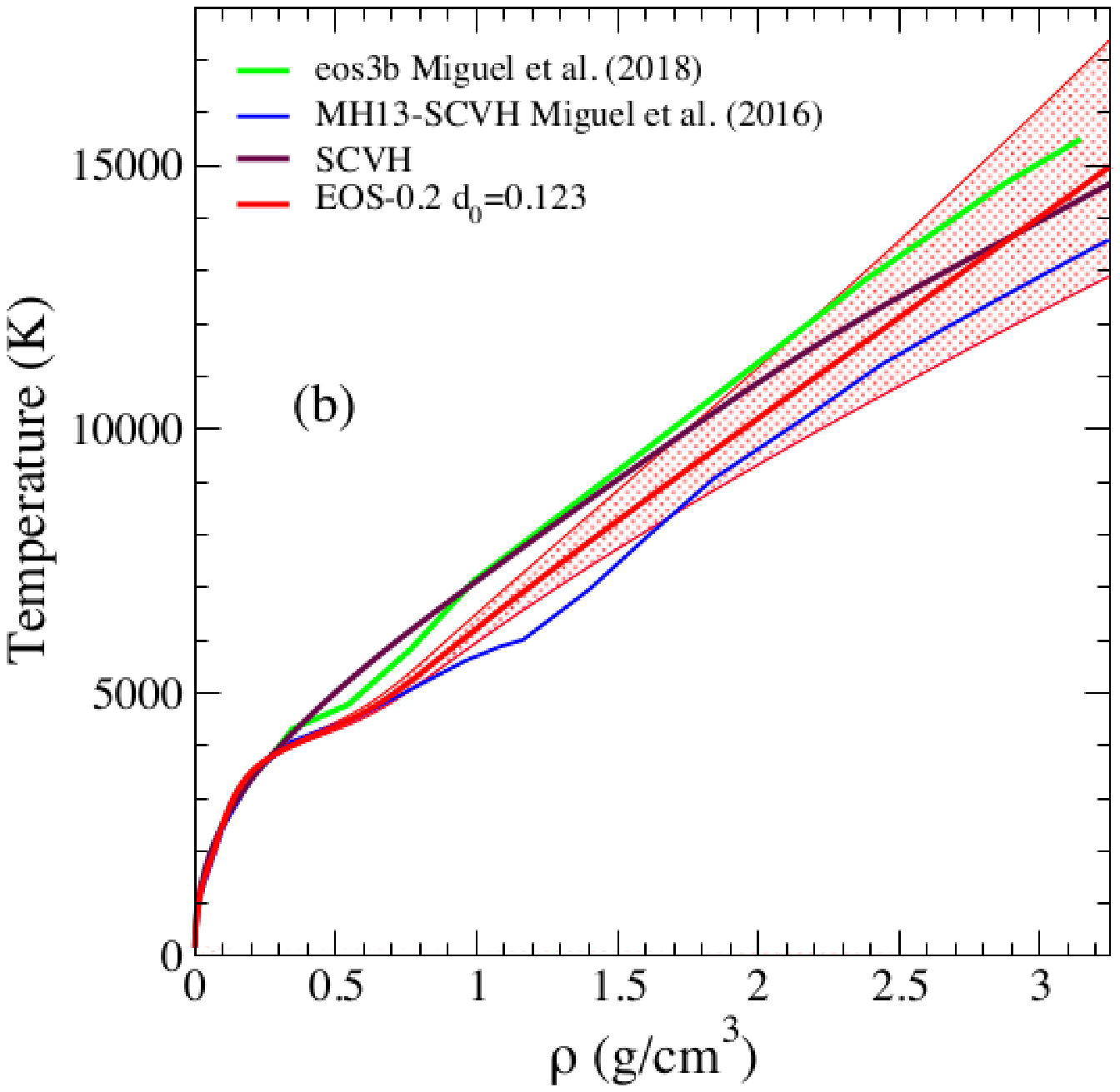}
   \caption{Jupiter interior profiles obtained with various version of the EOS adjusted at low and high densities and an helium mass fraction $Y_{He}=0.245$. (a) Comparison with previous {\sl ab initio} predictions of \citet{Militzer_2013}, \citet{Nettelmann_etal12}, and \citet{chabrier_2019}. (b) Comparison with the predictions of \citet{Miguel_Guillot_Fayon16} corrected by \citet{Miguel_2018}. The shaded area corresponds to the result obtained when d$_0$ is varied as in indicated in (a).} 
              \label{fig6}%
\end{figure}

We now turn to the predictions of the interior profile of Jupiter obtained using the various EOSs discussed in the previous section. We show in figure \ref{fig6} the
hydrogen-helium adiabat obtained using our recent EOS \citep{chabrier_2019} with an helium concentration $Y_{He}=0.245$. This profile is obtained by considering the planet as constituted
of an isentropic hydrogen-helium envelope with an homogeneous He concentration and a central core made of heavier elements. We used the temperature measured at $1~$bar by the Galileo
probe \citep{galileo}, $T_{1bar}=167~$K, to fixed the entropy of the isentrope using the SCVH EOS. This model assumes that the envelope is fully convective, neglects a potentially
radiative outer layer \citep{guillot_99}, the effect of demixing and of the multiple molecular species detected in the atmosphere by the Galileo probe \citep{hubbard_2016}.
With these effects neglected, the Jupiter's interior profiles calculated thus corresponds to the H-He isentrope.            

In figure \ref{fig6}-a, we compare our calculation for the H-He isentrope obtained for a fixed concentration of helium, $Y_{He}=0.245$, with the {\sl ab initio} predictions
of \citet{Militzer_2013} and \citet{Nettelmann_etal12}.  At low pressures, figure \ref{fig6}-a shows that varying the interpolation domain from 0.2 to 0.8 g/cm$^3$, noted respectively
EOS-0.2 and EOS-0.8, leads to a cooler isentrope in the $10-100~$GPa range. It also removes the change of slope noticeable from $10$ to $80~$GPa.  This suggests that a different
  interpolation scheme in this density region is likely at the origin of the difference obtained for the adiabat in the $10-300~$GPa range. This range also corresponds to
  tighter constrains from the Hugoniot data. We further see that the revision of the Hugoniot data confirms a slight variation of the slope of the adiabat of Jupiter as the
  pressure increase and dissociation takes place. We see that up to $100~$GPa, the agreement  with the prediction of \citet{Militzer_2013} is almost perfect while some departure is noticeable
with the calculation of \citet{Nettelmann_etal12}.  The latter is consistent with the differences noticed previously in the interpolation region between the SCVH and {\sl ab initio} data
  for, respectively, the internal energy  and the entropy. We further see that these variations propagate up to a few hundreds GPa in the isentrope.

   As pointed out previously \citep{Militzer_2013,Miguel_Guillot_Fayon16}, we see  a significant departure between the previous {\sl ab initio} predictions for pressures beyond $100~$GPa. The
  differences in temperature reaches up to $3000~$K at the core mantle boundary occurring at around $4000~$GPa. To identify the source of this discrepancy, we show the isentropes obtained by
  varying the $d_0$ parameters in the original fit of \citet{Caillabet_Mazevet_Loubeyre11}. The original value, $d_0=0.0326858$, and also corresponding to \citet{chabrier_2019}, is varied to
  $d_0=0.123$, to match exactly the initial {\sl ab initio} data, and  to $d_0=0.223$ to match the pressure data published by  \citet{Becker_etal14}. While this variation remains within the
  bounds of a fit designed for a large density-temperature range, we see that the impact for the isentrope of Jupiter is rather significant. Figure \ref{fig6}-a shows that the intermediate value
  for $d_0$ reproduces almost exactly the isentrope calculated by \citet{Militzer_2013}. We see that for $d_0=0.223$, which allows us to reproduce pressures obtained
  by \citet{Becker_etal14}, the high-pressure behavior of the isentrope obtained  by  \cite{Nettelmann_etal12}  is reproduced. While this isentrope is calculated with an earlier version of
  the hydrogen-helium equation of state \citep{holst_2008}, this comparison gives a plausible explanation for the ongoing discrepancies between these two {\sl ab initio}  predictions of the temperature at the core-envelope boundary.

\begin{figure}
   \centering
   \includegraphics[width=8cm]{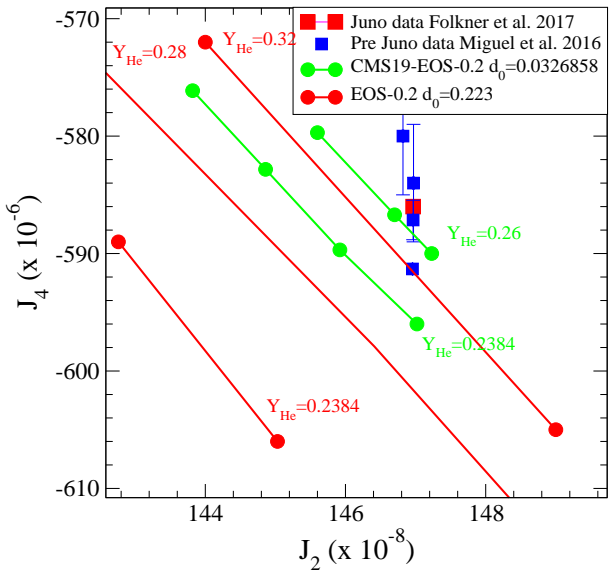}
   \caption{Variation of the gravitational moments, $J_2$ and $J_4$, as a function of the helium fraction, $Y_{He}$ calculated using a two-layer model, and EOSs with the two extreme values of $d_0$.}
   \label{fig7}
    \end{figure}

    In figure \ref{fig6}-b, we compare our EOSs obtained for different values of the $d_0$ parameter with the calculations of \citet{Miguel_Guillot_Fayon16}. We see that the trend
    noticed in figure \ref{fig6}-a remains the same.  More specifically, we see that the isentrope deduced from the \cite{Becker_etal14} data, noted eos3b, is consistently higher
      in temperature then the other EOSs in the $0.03-1.3~$g/cm$^3$ density range. It is even close to the SCVH isentrope in this density range. In contrast to \cite{Miguel_Guillot_Fayon16},
      we attribute this difference to the interpolation region between the SCVH and {\sl ab initio} data and the internal energy obtained rather than the method used
      to evaluate the entropy. Beyond a density of $1.5~$g/cm$^3$, we also see that the isentrope obtained reached higher temperatures when calculated using the \citet{Becker_etal14} EOS.
    The latter clips our estimation corresponding to the highest value of $d_0$.
    In contrast, the isentrope obtained  using a modified version of \citet{Militzer_2013} remains consistently colder. Figure \ref{fig6}-b shows that the mean value
    $d_0$  gives an isentrope that lays almost in between the interior profiles obtained using the \citet{Becker_etal14} EOS or based on the \citet{Militzer_2013} EOS. This result suggests
    that the MH13-SCVH result is not completely consistent with the initial  \citet{Militzer_2013} EOS calculated at a fixed helium concentration. We recall here that
    \citet{Miguel_Guillot_Fayon16} extract a pure hydrogen EOS by using the SCVH EOS for helium. As this does not coincide with the {\sl ab initio} results for pure
    helium \citep{soubiran2012}, it is not surprising that some differences remain for the isentrope. The comparison shown in figure \ref{fig7} suggests that the error is
    on the order of $1000~$K when using the EOS deduced by \citet{Miguel_Guillot_Fayon16}.

    We show, in figure \ref{fig7}, the variation of the first two gravitational moments obtained using the two extreme
    values of $d_0$. These calculations are performed using the theory of figures to the third order and considering a two-layers model consisting of an H-He envelope and a core made of water.
    While it is now well documented that such simple model is not sufficient to calculate the gravitational moments in the Juno era \citep{hubbard_2016,nettelmann_2017}, it enables us
    to quantify the uncertainty stemming from the EOS without having to consider the additional
    planetary model assumptions that vary from authors to authors. Following \citet{kerley_2013}, we propose that this model being used to benchmark future EOS intended for the modeling of Jupiter
    and the interpretation of the Juno data so as to decipher the uncertainties coming from the EOS from the ones arising from the planetary model.
    In figure \ref{fig7}, each curve represents an interior structure calculation at a fixed value of the helium
    concentration in the envelope while the size of the core varies. We assumed the core as made of pure water and used the {\sl ab initio} EOS for dense water \citep{mazevet_18} to
    model it. Figure \ref{fig7} shows that a larger amount of metallic elements is needed in the envelope to approach the measured values of the first two gravitational moments when
    using the EOS modified to match the \citet{Becker_etal14} EOS (EOS-0.2-$d_0=0.223$).  As pointed out by \citet{Wahl_etal17}, the size of the core does not change drastically
      when using the two different versions of the EOS corresponding to values of $d_0=0.023$ and $d_0=0.223$.
    
\begin{figure}
   \includegraphics[width=8cm]{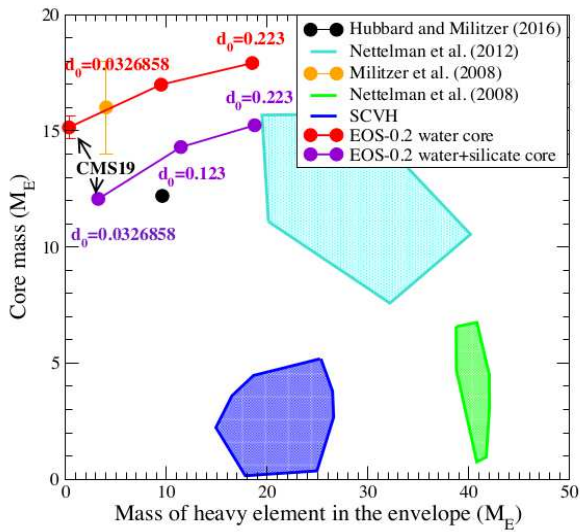}
   \caption{Predictions for the mass of the core and the amount of metallic elements in the envelope obtained using the 3rd order theory of figure. The published results are compared with the ones obtained by varying the $d_0$ parameter are shown in the figure and for a core constituted of pure water and of silicate and water.}
              \label{fig8}%
\end{figure}

This result is summarized in figure \ref{fig8}, where we compare the various predictions regarding the size of the core and the amount of metallic elements in the envelope as obtained
when we vary the value of the $d_0$ parameter to match previous {\sl ab initio} EOSs. We assume a core either made of pure water and described by the dense water
{\sl ab initio} EOS or a core made of water and silicates and described by the simple adiabatic pressure-density relations of \citet{hubbard_1989}.  Figure \ref{fig8} shows that
  the variation in the hydrogen {\sl ab initio} EOS, embodied in the value of $d_0$, leads to a variation in the amount of metallic elements predicted that is
  comparable to the one found by  \citet{Militzer_2013} and \citet{Nettelmann_etal12}. Based on the comparison shown before for the EOS and the isentrope, we suggest that the
  remaining difference with the \citet{Nettelmann_etal12} comes from the remaining difference for the isentrope around $1~$g/cm$^3$ that is not completely covered by varying the $d_0$ parameter.
  We further see that this result does not depend on the nature of the core as the amount of metallic elements predicted  in the envelope as $d_0$ is varied is similar for a pure water or a
  water-silicate core.

Our preferred EOS,  EOS-0.2 d$_0$=0.123, that coincides with the {\sl ab initio} simulation points to better than 1\%, and corrects the original \citet{Caillabet_Mazevet_Loubeyre11} and
\cite{chabrier_2019} EOS, is in rather good agreement with the predictions of \citet{Militzer_2013}.  The detailed study performed here shows that {\sl ab initio} EOS for hydrogen leads
  to Jupiter model with a low amount of metallic elements in the envelope. It furthermore provides an explanation for the differences reported with the predictions of
\citet{Nettelmann_etal12} that overestimates the amount of metallic elements in Jupiter's envelope. We further point out that this difference is likely due to the uncertainty in the
evaluation of the pressure either coming from the fit and/or the initial {\sl ab initio} data but not due to differences in the evaluation of the entropy as suggested by
\citet{Miguel_Guillot_Fayon16}. The latter only appears in the interpolation
region between the SCVH and {\sl ab initio} data  and, even in this case, is a consequence of the interpolation in energy and pressure between the {\sl ab initio} and SCVH data.

\section{Summary}
         Using our newly developed equations of state for hydrogen and helium \cite{chabrier_2019}, we investigate the long standing disagreement regarding the predictions of the amount of
          metallic elements in the envelope, the size of the core, and the temperature at the core-envelope boundary for Jupiter. We find the origin of the disagreement between the previous {\sl ab initio}
        predictions by varying the parameters in our parameterization. We confirm the prediction of \citet{Militzer_2013} and point out to deficiencies in the parameterization
        of  \citet{Becker_etal14} regarding the size of the core and the amount of metallic elements in the envelope. We also find that Jupiter inner structure and the associated
        gravitational moments are very sensitive to the evaluation of pressure and internal energy to a level that approaches the uncertainty in the {\sl ab initio} simulations. It further enters the regime where the different functionals used can have some influence \citep{Redmer_2018,Mazzola_2018}. This is a
        source of concern as neither the input {\sl ab initio} points or the fit developed for planetary modeling are brought to this level of accuracy. This result should be accounted
        for in more refined planetary models required to interpret the Juno data by using benchmarked equation of states carefully validated for all the quantities involved in planetary
        modeling, and particularly, the pressure, internal energy, and entropy. We suggest that a simple two-layers model provides a useful framework to benchmark future EOS. Application
        of this new EOS for hydrogen in a more refined planetary model as needed to interpret Juno data will be the object of further work.
        The benchmarked EOS for hydrogen, matching the initial {\sl ab initio} points of \citet{Caillabet_Mazevet_Loubeyre11} by less than 1\%, and providing an updated version of the CMS19 EOS,
        is provided at the following address https://luth.obspm.fr/\~luthier/mazevet/wordpress/planets-and-exoplanets/.
 
   \begin{acknowledgements}
     This work was funded by Paris Sciences et Lettres (PSL) university through the project origins and conditions for the emergence of life.  FS was supported by the European commission under the Marie Sklodowska-Curie project ABISSE -- grant agreement 750901. The authors also thanks T. Guillot for providing
     version of the EOSs produced in \cite{Miguel_2018} prior to publication and fruitful exchanges with M. Knudson regarding the analysis of the deuterium shock data.
\end{acknowledgements}

%
   \bibliographystyle{aa} 
   \bibliography{h2o} 
   \appendix
   \section{EOS and references}
   \begin{table}[ht]
     \begin{center}
       \begin{tabular}{|c|c|c|c|c|}
         \hline
         Name &  interpolation region in $\rho$ g/cm$ˆ$3 & d$_0$ & ab initio data& reference \\
         \hline
         CMS19-EOS-0.2 d$_0$=0.0326858 & 0.05-0.2 & 0.0326858 & \cite{Caillabet_Mazevet_Loubeyre11}&\cite{chabrier_2019}\\
         \hline
         EOS-0.8 d$_0$=0.0326858 &0.05-0.8& 0.0326858&\cite{Caillabet_Mazevet_Loubeyre11}&\\
         \hline
         EOS-0.2 d$_0$=0.123 &0.05-0.2& 0.123& \cite{Caillabet_Mazevet_Loubeyre11}&This work\\
         \hline
         EOS-0.2 d$_0$=0.223 & 0.05-0.2 & 0.223 &\cite{Caillabet_Mazevet_Loubeyre11}&\\
         \hline
         eos3b &-&-&\cite{Becker_etal14}&\cite{Miguel_2018}\\
         \hline
         eos3c &-&-&\cite{Becker_etal14}&\cite{Miguel_Guillot_Fayon16}\\
         \hline
         MH-SCVH&-&-&\cite{Militzer_2013}&\cite{Miguel_Guillot_Fayon16}\\
         \hline
       \end{tabular}
     \end{center}
     \end{table}
         
   \end{document}